\documentclass[usenatbib]{mn2e}
\usepackage{graphicx}

\title[Ultraviolet Fe\,{\normalsize\it II} emission in $z \sim 2$ quasars]
{Ultraviolet Fe\,{\Large\bf II} emission in $\bmath{z \sim 2}$ quasars}
\author[H. Sameshima]{H. Sameshima$^1$\thanks{E-mail: hsameshima@ioa.s.u-tokyo.ac.jp}, J. Maza$^2$,
Y. Matsuoka$^1$, S. Oyabu$^3$, K. Kawara$^1$, Y. Yoshii$^1$, \newauthor
N. Asami$^1$, N. Ienaka$^1$ and Y. Tsuzuki$^4$
\\
$^1$Institute of Astronomy, University of Tokyo, 2-21-1, Osawa, Mitaka, Tokyo 181-0015, Japan \\
$^2$Departamento de Astronomia, Universidad de Chile, Casilla 36-D, Santiago, Chile \\
$^3$Institute of Space and Astronautical Science, Japan Aerospace Exploration Agency, Kanagawa 229-8510, Japan \\
$^4$System Integration Service Department, Redfox Inc., 3-3-11, Kita-Aoyama, Minato-ku, Tokyo 107-0061, Japan}

\begin{document}

\maketitle

\begin{abstract}
  
  We present spectra of six luminous quasars at $z \sim 2$, covering rest wavelengths 
  1600$-$3200 \AA. The fluxes of the UV Fe\,{\sevensize II} emission lines and 
  Mg\,{\sevensize II} $\lambda 2798$ doublet, the line widths of Mg\,{\sevensize II}, 
  and the 3000 \AA\ luminosity were obtained from the spectra. These quantities were 
  compared with those of low-redshift quasars at $z = 0.06 - 0.55$ studied by Tsuzuki 
  et al. In a plot of the Fe\,{\sevensize II}(UV)/Mg\,{\sevensize II} flux ratio as a 
  function of the cental black hole mass, Fe\,{\sevensize II}(UV)/Mg\,{\sevensize II} 
  in our $z \sim 2$ quasars is systematically greater than in the low-redshift quasars. 
  We confermed that luminosity is not responsible for this excess. It is unclear whether 
  this excess is caused by rich Fe abundance at $z \sim 2$ over low-redshift or by 
  non-abundance effects such as high gas density, strong radiation field, and high 
  microturbulent velocity.   

\end{abstract}


\section{Introduction}

According to the models of explosive nucleosynthesis, much of the iron comes from Type Ia supernovae, 
while $\alpha$ elements such as O and Mg come from Type II supernovae. Because the difference in 
lifetime of the progenitors, it is generally considered that the iron enrichment delays relative to 
$\alpha$ elements by 1--2 billion years (\citealt{hf}; \citealt{ysi2}, 1998).
If Fe\,{\sevensize II}/Mg\,{\sevensize II}, the relative strengths of Fe\,{\sevensize II} emission lines 
and the Mg\,{\sevensize II} $\lambda 2798$ doublet, reflects the Fe/Mg abundance ratio, there will be 
a break in Fe\,{\sevensize II}/Mg\,{\sevensize II} at high redshift. Despite of much efforts made by many 
observational groups (e.g., \citealt{els}; \citealt{kwr}; \citealt{die2} , 2003; \citealt{iwa2}, 2004; 
\citealt{fre}; \citealt{mai}; \citealt{tzk}; \citealt{mat2}, 2008a; \citealt{kur}), there have been found 
no signs of such a break; Fe\,{\sevensize II}/Mg\,{\sevensize II} looks constant from low-redshift up to 
$z \sim 6.5$ with large scatter. 

No break in Fe\,{\sevensize II}/Mg\,{\sevensize II} might reflect a significantly shorter delay-time 
of 0.2$-$0.6 Gyr, as suggested by \citet{ft}, \citet{mr}, and \citet{gra}. The expected break can also 
be obscured by non-abundance effects. Simulations of Fe\,{\sevensize II} emitting regions, assuming either 
photoionization or shocks, imply that the Fe abundance is not only one parameter which controls the 
Fe\,{\sevensize II} strength, but several non-abundance factors can also affect it. Such non-abundance 
factors include spectral energy distribution (SED) of the central source, strength of the radiation field, 
and the gas density of Broad Emission Line Region (BELR) clouds. Recently, \citet{ver} and \citet{bal} 
pointed out that a large microturbulence velocity may be responsible for strong Fe\,{\sevensize II} emission. 
\citet{tzk} have studied non-abundance factors by using spectra of a low-redshift sample of 14 quasars, 
covering wide rest-wavlengths 1000$-$7300 \AA, and claimed that the Fe\,{\sevensize II} strength correlates 
with the mass of the central black hole, the line width, and the X-ray photon index.

In this paper, we present spectra of six quasars at $z\sim2$, and compare with those in the low-redshift 
sample.  Throughout this paper, a cosmology with $\Omega_m=0.3,\ \Omega_{\Lambda}=0.7$, and 
$H_0=70\ \mathrm{km\ s^{-1}\ Mpc^{-1}}$ is assumed.

\section{Observations}
Six quasars were selected for optical spectroscopy from the catalog by \citet{vv}.
According to the catalog, these are luminous with $M_B=-28$ -- $-31$ at $z=2.0$ -- $2.3$ , bright enough to 
take optical Fe\,{\sevensize II} emission lines through near-infrared spectroscopy at a later opportunity.  

GMOS on Gemini-South Telescope are used in the long-slit mode with grating R150\_G5326 and order sorting 
filter OG515\_G0330. Wavelengths observed are in 5400$-$9800 \AA, corresponding to rest wavelengths 
1800$-$3300 \AA\ at $z\sim2$ quasars. The slit width is 1.0'' and the spectral resolution is 3.286 \AA\ 
pixel$^{-1}$. The grating was centered at 8150 \AA\ for the first three exposures and changed to 8250 \AA\ 
for the following exposures, filling up the gaps of the CCD chip array. Wavelengths were calibrated using 
the CuAr arc lamp taken at the both central wavelengths of the grating. LTT1788 (\citealt{ham}) was used 
for flux-scaling. The observing log is summarlized in Table \ref{tab:obslog}.

\begin{figure}
  \includegraphics[width=84mm]{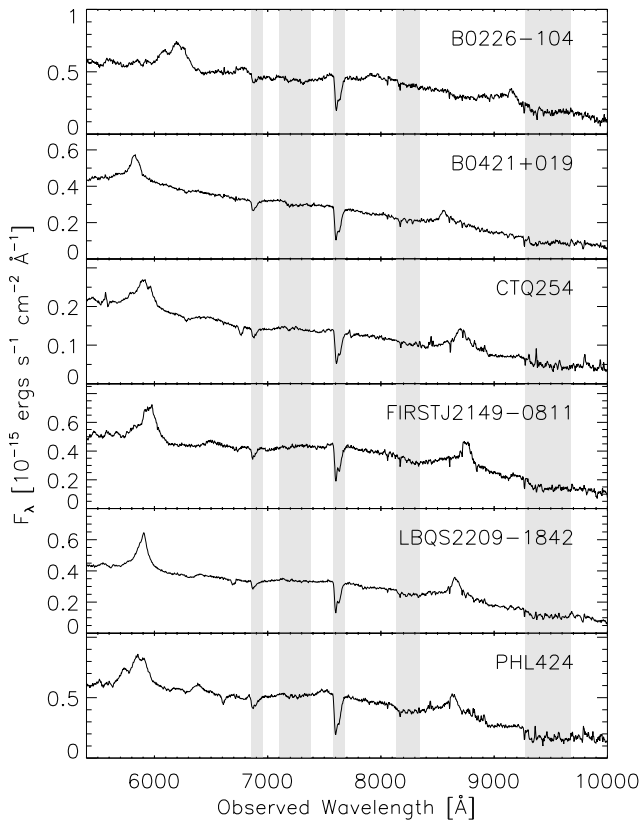}
  \caption{
    Spectra of six quasars.
    The shaded area indicates the region where the spectra are affected by the telluric absorption.
  }
  \label{fig:spec}
\end{figure}

Individual spectral frames were processed using the Gemini IRAF package ver1.9.1. Sky of individual 
frames is subtracted with GSSKYSUB, and then SCOMBINE was used to combine into the final spectral frame. 
The corrections for telluric absorption were not applied. Reduced spectra are shown in Figure \ref{fig:spec}.
Redshifts are determined from fit of Mg\,{\sevensize II} emission line. With these measured redshifts, 
absolute $B$ magnitudes are calculated assuming a cosmology with $\Omega_m=0.3,\ \Omega_{\Lambda}=0.7$, and 
$H_0=70\ \mathrm{km\ s^{-1}\ Mpc^{-1}}$. Redshifts and absolute $B$ magnitudes are also listed in Table \ref{tab:obslog}.

\begin{table*}
  \centering
  \begin{minipage}{155mm}
    \caption{Observing Log for Gemini Quasars\label{tab:obslog}}
    \begin{tabular}{lccccccc}
      \hline \hline
      Object & $\alpha$ & $\delta$ & Redshift\footnote{Redshift as determined from fit of 
      Mg\,{\sevensize II} emission line.} & $M_B$\footnote{Absolute $B$ magnitude with 
        $\Omega_m=0.3,\ \Omega_{\Lambda}=0.7$, and $H_0=70\ \mathrm{km\ s^{-1}\ Mpc^{-1}}$.}
      & $E_{B-V}$\footnote{Galactic extinction $E_{B-V}$ taken from \citet{sch}} & Exposure Time & Date \\
       & & & & [mag] & [mag] & [s] & \\
      \hline
      B0226$-$104       & 02 28 39.2 & $-$10 11 10 & 2.276 & $-29.7$ & 0.03 & 120  & 2004 Sep 18 \\
      B0421+019         & 04 24 08.6 &   +02 04 25 & 2.059 & $-27.8$ & 0.19 & 600  & 2004 Sep 18 \\
      CTQ254            & 04 30 14.6 & $-$36 26 47 & 2.118 & $-27.7$ & 0.02 & 1500 & 2004 Sep 18 \\
      FIRSTJ2149$-$0811 & 21 49 48.2 & $-$08 11 16 & 2.128 & $-28.9$ & 0.04 & 150  & 2004 Sep 20 \\
      LBQS2209$-$1842   & 22 12 10.4 & $-$18 27 38 & 2.093 & $-27.4$ & 0.03 & 960  & 2004 Sep 20 \\
      PHL424            & 23 13 24.5 &   +00 34 45 & 2.087 & $-28.5$ & 0.04 & 300  & 2004 Sep 21 \\
      \hline
    \end{tabular}
  \end{minipage}
\end{table*}

\section{Measurement of emission lines} \label{sec:measure}
Prior to measuring physical quantities such as Fe\,{\sevensize II} emission lines, the quasar spectra 
were dereddened for the Galactic extinction according to the dust map by \citet{sch} using the Milky 
Way extinction curve by \citet{pei}. $E_{B-V}$ of the Galactic extinction is listed in Table 
\ref{tab:obslog}.  In the shaded area in Figure \ref{fig:spec}, the telluric absorption features are 
seen. We have not applied any correction for the telluric absorption. Instead, the intensities in the 
shaded areas were estimated by fitting a linear function to assumed data points locating on either 
side free from the telluric features.

\subsection{Fe\,{\sevensize II} UV emission lines}

Fe\,{\sevensize II} emission lines are heavily blended with each other, forming the broad features 
from 2000$-$3000 \AA. It is desirable to observe a wide range of wavelengths in such a way that the 
power-law and Balmer continua are accurately determined as made by \citet{tzk}. However, observing 
such a wide range is not feasible in most cases. In fact, the present observations are limited to 
a rest wavelength range from 1600$-$3200 \AA.  
 
\begin{figure}
  \includegraphics[width=84mm]{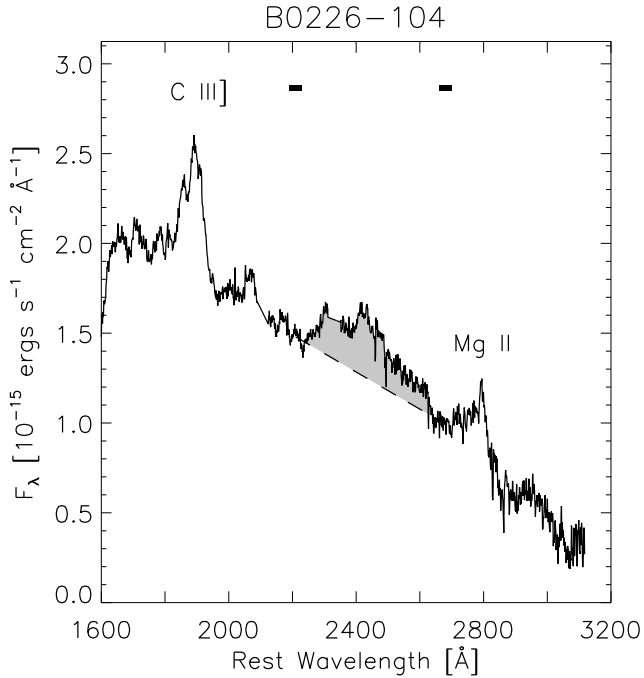}
  \caption{
    Measuring the UV Fe\,{\sevensize II} line flux.
    A linear fuction is fit to the data in the continuum windows 
    ({\it thick bars}). The best-fit function is indicated by the dashed line.
    The shaded area indicate $f(2240-2650\ \mathrm{\AA})$, which is then converted to 
    {\it Fe\,{\sevensize II}}\,$(2000-3000\ \mathrm{\AA})$ using equation (\ref{eq:feiiconv}).
  }
\label{fig:feii}
\end{figure}

\begin{figure}
  \includegraphics[width=84mm]{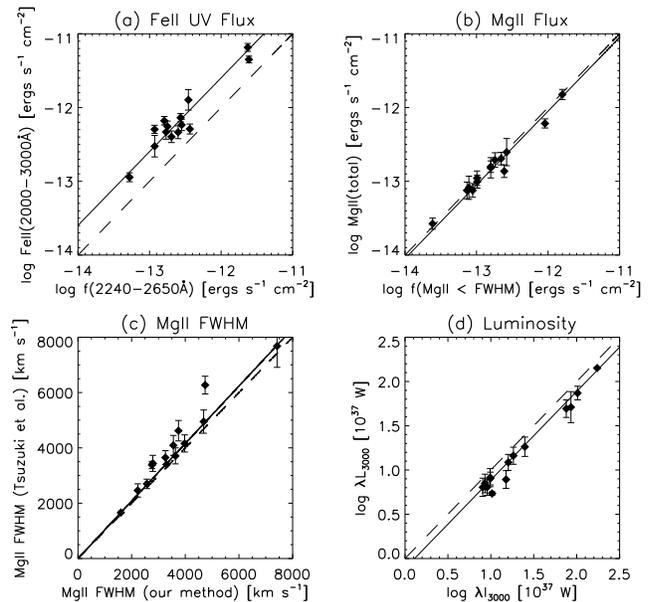}
  \caption{
    Comparison of the physical quantities in the low-redshift sample measured between the two methods; 
    data along the {\it horizontal}-axis are measured by our method and those along the 
    {\it vertiacal}-axis by \citet{tzk}. The dashed line indicate $y=x$ where two measuements result 
    in the identical values, and the solid line indicates the best-fit line. Each panel shows
    $(a)$ Fe\,{\sevensize II}(2000$-$3000 \AA) [$\mathrm{ergs\ s^{-1}\ cm^{-2}}$], 
    $(b)$ Mg\,{\sevensize II}({\it total}) [$\mathrm{ergs\ s^{-1}\ cm^{-2}}$], 
    $(c)$ FWHM(Mg\,{\sevensize II}) [$\mathrm{km\ s^{-1}}$], 
    $(d)$ 3000 \AA\ Luminosity [$10^{37}\ \mathrm{W}$].
  }
  \label{fig:corr}
\end{figure}

We applied a simple alternative in which a linear function is fit to the data in rest wavelengths 
2190$-$2230 \AA\ and 2660$-$2700 \AA. Differences between the spectrum and the resultant best-fit 
function, which are marked as shade in Figure \ref{fig:feii}, are summed up in a wavelengh range 
of 2240$-$2650 \AA. The summed-up differences, as denoted by $f(2240-2650\ \mathrm{\AA})$, should 
contribute siginificant part of {\it Fe\,{\sevensize II}}\,$(2000-3000\ \mathrm{\AA})$ which is 
the {\it total} Fe\,{\sevensize II} emission line flux in 2000$-$3000 \AA. To check the relationship 
between them, we have applied this alternative to the low-redshift quasars studied by \citet{tzk} for 
which {\it Fe\,{\sevensize II}}\,$(2000-3000\ \mathrm{\AA})$ is known. The results are shown in 
Figure \ref{fig:corr}$(a)$. This figure indicates that the relation is linear and 
$f(2240-2650\ \mathrm{\AA})$ is approximately 40\% of {\it Fe\,{\sevensize II}}\,$(2000-3000\ \mathrm{\AA})$.
A least-squares fitting to the data gives the following relation:

\begin{equation}
  \log \,${\it Fe\,{\sevensize II}}$\, = \log f + 0.402 (\pm 0.142) \label{eq:feiiconv}
\end{equation}
Here {\it Fe\,{\sevensize II}} $\equiv$ {\it Fe\,{\sevensize II}}\,$(2000-3000\ \mathrm{\AA})$ 
and $f \equiv f(2240-2650\ \mathrm{\AA})$.

This relation will be used to convert observed $f(2240-2650\ \mathrm{\AA})$ to the total flux of the 
Fe\,{\sevensize II} emission lines in 2000$-$3000 \AA\ in the later part of this paper.

\subsection{Mg\,{\sevensize II} emssion lines}

To measure the flux and the full width at half maximum (FWHM) of the Mg\,{\sevensize II} $\lambda 2798$ doublet, 
\citet{tzk} fitted a single Gaussian component to the spectrum where the power-low and Balmer continua, and 
Fe\,{\sevensize II} emission features were already subtracted. Again, we are not allowed to apply their method 
because of our limited wavelength range. 

\begin{figure}
  \includegraphics[width=84mm]{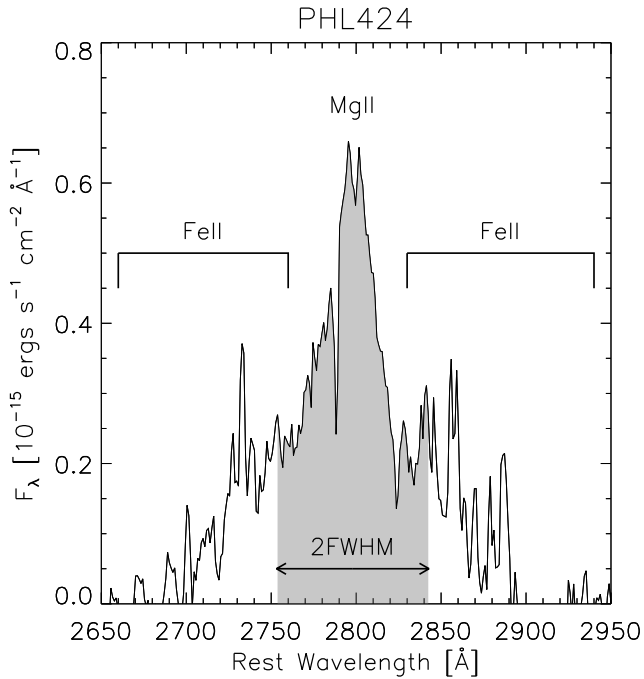}
  \caption{
    Measuring the Mg\,{\sevensize II} flux.
    Since there are Fe\,{\sevensize II} emissions underneath Mg\,{\sevensize II}, we measured the flux 
    only over the velocity range $-\mathrm{FWHM} < v(\mathrm{Mg\ II}) < \mathrm{FWHM}$ and defined it 
    as $f$(Mg\,{\sevensize II} $<$ FWHM) (shaded area).
  }
  \label{fig:mgii}
\end{figure}

Our alternative is illustrated in Figure \ref{fig:mgii}. A linear function is fit to the data in 2660$-$2700 
\AA\ and 2930$-$2970 \AA\ where contributions from the Fe\,{\sevensize II} 
and Mg\,{\sevensize II} emission lines are relatively weak and thus the power-law continuum can be 
defined\footnote{3000$-$3050 \AA\ would be a better choice than 2930$-$2970 \AA\ if the CCD fringe 
pattern can be well removed in these wavelenghts.}. This fitted function is subtracted from the spectrum, 
as shown in Figure \ref{fig:mgii}. To measure the Mg\,{\sevensize II} FWHM, we first applied smoothing to that 
subtracted spectrum, then measured the velocity range within which the flux become more than half of its maximum 
value and defined it as FWHM. To minimize contributions from Fe\,{\sevensize II} emission, we only integrate the flux 
within $-\mathrm{FWHM} < v(\mathrm{Mg\ II}) < \mathrm{FWHM}$ and define it as $f($Mg\,{\sevensize II} $<$ FWHM$)$. 
Again, we used the low-redshift sample by \citet{tzk} to check that the relation between 
$f($Mg\,{\sevensize II} $<$ FWHM$)$ and the {\it total} Mg\,{\sevensize II} flux {\it Mg\,{\sevensize II}(total)}. 
The results are shown in Figure \ref{fig:corr}$(b)-(c)$.  $f($Mg\,{\sevensize II} $<$ FWHM$)$ has linear relation to the 
{\it total} Mg\,{\sevensize II} line flux measured by \citet{tzk}, and the least-squares best-fit to the 
data gives the following relation:

\begin{equation}
  \log \,${\it Mg\,{\sevensize II}(total)}$\, = \log f($Mg\,{\sevensize II} $ < $ FWHM$) - 0.049 (\pm 0.084) \label{eq:mgiiconv}
\end{equation}

This will be used to obtain {\it Mg\,{\sevensize II}(total)} in the later part of this paper. It is noted 
that there are no significant differences in FWHM of Mg\,{\sevensize II} between the measurements by 
\citet{tzk} and our alternative.

\subsection{Luminosity}
\citet{mj} gives a method for estimating black hole masses of quasars using the FWHM of the 
Mg\,{\sevensize II} emission line and the continuum luminosity at 3000 \AA. The equation is as follows:
\begin{equation}
  \frac{M_{\mathrm{BH}}}{M_{\odot}}=3.37 \left( \frac{\lambda L_{3000}}{10^{37}\mathrm{W}} \right)^{0.47} 
  \left[ \frac{\mathrm{FWHM(Mg\ II)}}{\mathrm{km\ s^{-1}}} \right]^2
  \label{eq:mclure}
\end{equation}
Note that the uncertainty of equation (\ref{eq:mclure}) is a factor of 2.5.

Hence the monochromatic luminosity $\lambda L_{3000}$ has to be measured in order to estimate the black 
hole mass. Unfortunately, however, our spectra are significantly affected by telluric absorption at 3000 
\AA. We thus extrapolated the linear function, which was used to measure $f($Mg\,{\sevensize II} $<$ FWHM$)$, 
to 3000 \AA\, and read the value at 3000 \AA\ as a monochromatic flux. Figure \ref{fig:corr}$(d)$ compares 
the 3000 \AA\ luminosity $\lambda L_{3000}$ measured by \citet{tzk} and our extrapolated 3000 \AA\ luminosity 
$\lambda l_{3000}$. Again, the linear relation was obtained as follows:

\begin{equation}
  \log \lambda L_{3000}= \log \lambda l_{3000} - 0.112 (\pm 0.079) \label{eq:lumiconv}
\end{equation}

This will be used to convert extrapolated luminosities to 
the real luminosity in the later part of this paper.

\begin{table*}
  \centering
  \begin{minipage}{170mm}
    \caption{Measured Physical Quantities \label{tab:measure}}
    \begin{tabular}{lccccc}
      \hline \hline
      Object & {\it Fe\,{\sevensize II}}\,(2000$-$3000 \AA)\footnote{Fe\,{\sevensize II} is difined in a wavelength range of 2000$-$3000 \AA.} 
             & {\it Mg\,{\sevensize II}(total)} & FWHM(Mg\,{\sevensize II})\footnote{FWHM of the Mg\,{\sevensize II} emission line.} & $\lambda L_{3000}$ & $M_{\mathrm{BH}}$ \\
             & [$10^{-14}\ \mathrm{ergs\ s^{-1}\ cm^{-2}}$] & [$10^{-14}\ \mathrm{ergs\ s^{-1}\ cm^{-2}}$] & [km/s] & [$10^{37}\ \mathrm{W}$] & [$10^{9}\ M_\odot$] \\
      \hline
      B0226$-$104            & 19.3 (+7.5/$-$5.4) & --- & ---   & ---  & ---   \\
      B0421+019              & 5.46 (+2.1/$-$1.5)  & 1.24 (+0.26/$-$0.22)  & 4410 (+520/$-$470) & 369 (+74/$-$62)  & $1.05\pm0.269$  \\
      CTQ254                 & 3.11 (+1.2/$-$0.87) & 0.726 (+0.16/$-$0.13) & 4750 (+570/$-$510) & 161 (+32/$-$27)  & $0.828\pm0.212$ \\
      FIRSTJ2149$-$0811      & 14.1 (+5.5/$-$3.9)  & 2.16 (+0.46/$-$0.38)  & 3900 (+460/$-$410) & 500 (+100/$-$83) & $0.951\pm0.243$ \\
      LBQS2209$-$1842        & 8.78 (+3.4/$-$2.5)  & 1.70 (+0.36/$-$0.30)  & 4180 (+500/$-$440) & 360 (+72/$-$60)  & $0.935\pm0.239$ \\
      PHL424                 & 19.5 (+7.6/$-$5.5)  & 2.81 (+0.60/$-$0.49)  & 5000 (+590/$-$530) & 587 (+120/$-$98) & $1.69\pm0.431$  \\
      \hline
    \end{tabular}
  \end{minipage}
\end{table*}

\section{Results}

In Table \ref{tab:measure}, the Fe\,{\sevensize II} emission line flux 
{\it Fe\,{\sevensize II}}\,$(2000-3000\ \mathrm{\AA})$, the Mg\,{\sevensize II} line flux 
{\it Mg\,{\sevensize II}(total)}, the FWHM (Mg\,{\sevensize II}), the 3000 \AA\ luminosity $\lambda L_{3000}$, 
and the blackhole mass $M_{BH}$ derived from equation (\ref{eq:mclure}), are given. Note that, for B0226-104, 
Mg\,{\sevensize II} emission line is heavily affected by telluric absorption line, so that we did not measure 
Mg\,{\sevensize II} flux and Mg\,{\sevensize II} FWHM. A plot of 
{\it Fe\,{\sevensize II}}\,$(2000-3000\ \mathrm{\AA})$/{\it Mg\,{\sevensize II}(total)} against $M_{BH}$ is 
given in Figure \ref{fig:femg}$(a)$ and 3000 \AA\ luminosity $\lambda L_{3000}$ in Figure \ref{fig:femg}$(b)$.

\section{Discussion}

\begin{figure*}
  \includegraphics[width=160mm]{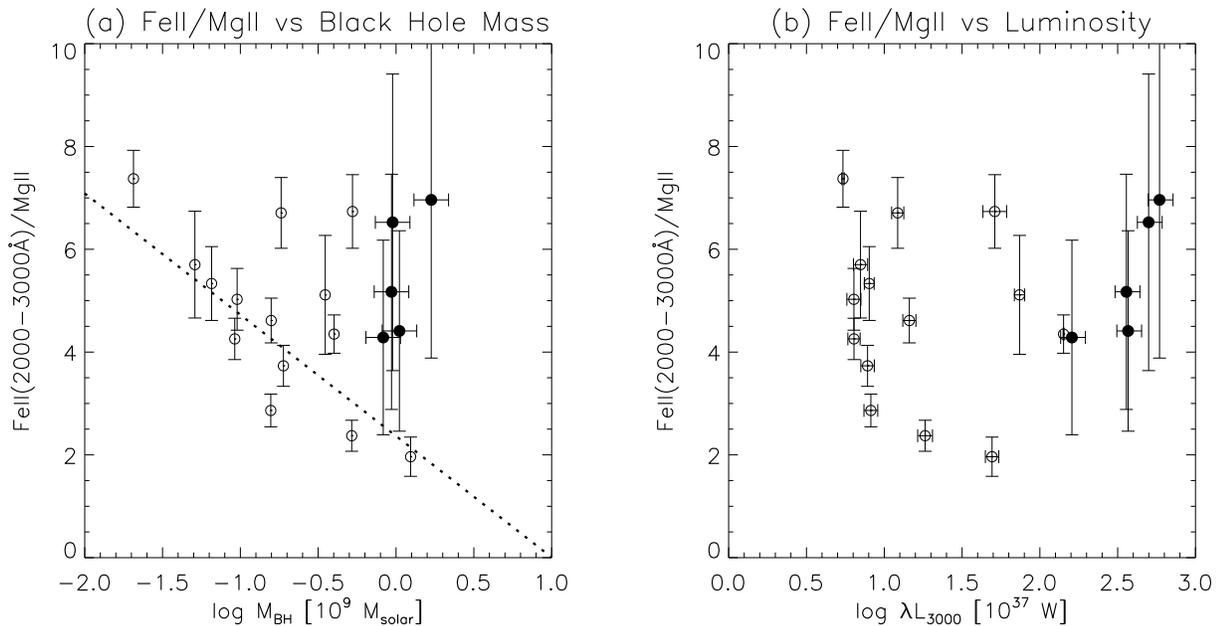}
  \caption{
    $(a)$ Relation between Fe\,{\sevensize II}(UV)/Mg\,{\sevensize II} flux ratio and black hole mass.
    Filled circles are our quasars at $z \sim 2.0$. Open circles are low-redshift quasars at $z = 0.06 - 0.55$ 
    by \citet{tzk}. Dotted line indicates the correlation found in \citet{tzk}. $(b)$ Same as in $(a)$, but for 
    3000 \AA\ luminosity.
  }
\label{fig:femg}
\end{figure*}

Analyzing 14 low-redshift quasars, \citet{tzk} found the correlation between the flux ratio 
Fe\,{\sevensize II}(UV)/Mg\,{\sevensize II} and the blackhole mass. This relation is shown by the dotted line 
in Figure \ref{fig:femg}$(a)$. Filled circles are our quasars at $z \sim 2.0$ and open circles are low-redshift 
quasars at $z = 0.06 - 0.55$ by \citet{tzk}. Our quasars have an absolute luminosity of $M_B < -27.4$, which 
are much luminous than the low-redshift quasars having an absolute luminosity of $M_B > -26.3$. 

As can be seen in Figure \ref{fig:femg}$(a)$, the five $z \sim 2$ quasars do not follow the correlation found 
by \citet{tzk}. All of them have Fe\,{\sevensize II}(UV)/Mg\,{\sevensize II} greater than expected from the 
Tsuzuki's correlation. What is the cause of the large Fe\,{\sevensize II}(UV)/Mg\,{\sevensize II} value in 
the $z \sim 2$ quasars relative to the low-redshift quasars? Because the five $z \sim 2$ quasars are much luminous 
than the 14 low-redshift quasars, the luminosity effect is examined. As shown in Figure \ref{fig:femg}$(b)$, 
the luminosity effect is not responsible for the large Fe\,{\sevensize II}(UV)/Mg\,{\sevensize II} value in the 
$z \sim 2$ quasars. The {\it real} cause would be evolution in Fe\,{\sevensize II}(UV)/Mg\,{\sevensize II} or 
non-abundance effects such as the sepectral energy distribution of the continuum from the central source,
the strength of the radiation field and the gas density of BLR clouds as well as the microturbulence of BLR 
gas (\citealt{ver}; \citealt{bal}). Further investigations are required using large samples of quasars.

\section*{acknowledgements}
This work was financially supported in part by Grant-in-Aid for Scientific Research (17104002) and Specially 
Promoted Reserch (20001003) and the Japan-Australia Research Cooperative Program from JSPS.

\end{document}